____________________________________________________________________________
\documentstyle[12pt]{article}
\title{Irreducible bases and correlations of spin states\\
for double point groups} 

\author{Shi-Hai Dong$^{1)}$, Xi-Wen Hou$^{1,2)}$, and Zhong-Qi Ma$^{1)}$\\
{\footnotesize  1) Institute of High Energy Physics, P.O.Box 918(4), Beijing 
100039, P. R. of China}\\
{\footnotesize  2) Department of Physics, University of Three Gorges, 
Yichang 443000, P. R. of China}}

\date{}
\begin{document}
\maketitle

\begin{abstract}
In terms of the irreducible bases of the group space of the
octahedral double group {\bf O'}, an analytic formula is 
obtained to combine the spin states $|j,\mu \rangle$ into 
the symmetrical adapted bases, belonging to a given row of 
a given irreducible representation of {\bf O'}. This method
is effective for all double point groups.  However, for the 
subgroups of {\bf O'}, there is another way to obtain those
combinations. As an example, the correlations of
spin states for the tetrahedral double group {\bf T'} are 
calculated explicitly. 
\end{abstract}

\newpage
\noindent
{\bf 1. Introduction}

\vspace{3mm}
It is a common problem to combine the spin states $|j,\mu \rangle$
into the symmetrical adapted bases (SAB), that are defined as the 
orthogonal bases belonging to the given rows of the given 
irreducible representations of a point group. The SAB are
very useful in classifying the electronic states in the present of
spin coupling. Especially, for the electronic states with 
half-odd-integer spin, one has to deal with the double group 
symmetry \cite{be}. 

As is well known, $SU(2)$ group is the covering group of
the rotation group $SO(3)$, and provides the double-valued
representations of $SO(3)$. Following the homomorphism of 
$SU(2)$ onto $SO(3)$:
$$\pm u(\hat{{\bf n}},\omega)~\longrightarrow ~R(\hat{{\bf n}},\omega).
\eqno (1) $$

\noindent
we are able to define the double point groups as follows. In the 
rotation group $SO(3)$, a rotation through $2\pi$ is equal to identity 
$E$, but it is different from identity in the $SU(2)$ group:
$$R(\hat{{\bf n}},2\pi)=E,~~~~~u(\hat{{\bf n}},2\pi)
\equiv E'=-{\bf 1}.  \eqno (2) $$

\noindent
The point group $G$ is a subgroup of $SO(3)$, and the double 
point group $G'$ is that of $SU(2)$. A point group $G$ is 
extended into a double point group $G'$ \cite{be} by introducing 
the new element $E'$, satisfying:
$$RE'=E'R,~~~~~(E')^{2}=E,~~~~~R\in G\subset G',~~~~E'R\in G'.
\eqno (3) $$

\noindent
In order to distinguish $R\in G \subset G'$ from $E'R \in G'$, we 
restrict the rotation angle $\omega$ not larger than $\pi$:
$$\begin{array}{l}
u(\hat{{\bf n}},\omega)~\longrightarrow~R(\hat{{\bf n}},\omega)\in G,
~~~~~~~~~0\leq \omega \leq \pi  , \\
-u(\hat{{\bf n}},\omega)=u(-\hat{{\bf n}},2\pi-\omega)~
\longrightarrow~R(-\hat{{\bf n}},2\pi-\omega)=R(\hat{{\bf n}},\omega-2\pi)
=E'R(\hat{{\bf n}},\omega). \end{array} \eqno (4) $$

\noindent
The period of $\omega$ in $SU(2)$ group is $4\pi$. 

Recently, the problem of combining the spin states $|j,\mu \rangle$
into the SAB has drawn some attention
of physicists. A new technique \cite{chen}, called the 
double-induced technique, was used for calculating the 
irreducible bases for the tetrahedral group {\bf T'}
and the combinations of the angular momentum states. It
was announced \cite{chen} that this technique will be 
used to calculate the similar problems for the octahedral 
double group {\bf O'} and icosahedral double group {\bf I'}.
The character table and the correlation tables relevant 
for the icosahedral double group {\bf I$_{h}'$} were 
presented apparently \cite{bal}. The element $E'$ was 
denoted by $R$ in \cite{bal} and \cite{ham}, and by $\theta$ 
in \cite{chen}. The double point group $G'$ was denoted by 
$G^{\dagger}$ in \cite{chen}.

As is well known, the character tables of double point 
groups are easy to be obtained from the group theory, 
and the calculation for the correlation tables is
straightforward. In principle, the character table can be 
used to find the similarity transformation that combines the 
states with a given angular momentum into the SAB. However, 
it becomes a tedious task while the angular momentum 
increases. Fortunately, the difficulty can be conquered 
by the irreducible bases in the group space.

From group theory (\cite{ham} p.106), the group space
is the representation space of the regular representation
where the natural bases are the group elements.
The number of times each irreducible representation is
contained in the regular representation is equal to the 
dimension of the representation. Reducing the regular 
representation, we can obtain the irreducible bases 
$\psi^{\Gamma}_{\mu \nu}$ with the following property:
$$R\psi^{\Gamma}_{\mu \nu}=\displaystyle \sum_{\rho}~
\psi^{\Gamma}_{\rho \nu}D^{\Gamma}_{\rho \mu}(R),~~~~~~
\psi^{\Gamma}_{\mu \nu}R=\displaystyle \sum_{\rho}~
D^{\Gamma}_{\nu \rho}(R)\psi^{\Gamma}_{\mu \rho}. 
 \eqno (5) $$

\noindent
Therefore, those irreducible bases are called the bases 
belonging to the $\mu$ row and the $\nu$ column of the 
irreducible representation $\Gamma$.

Assume that $G$ is a point group, which is a subgroup of 
the rotation group $SO(3)$. Applying its irreducible
bases $\psi^{\Gamma}_{\mu \nu}$ to the angular momentum 
states $|j, \rho \rangle$, we obtain the SAB 
$\psi^{\Gamma}_{\mu \nu}~|j, \rho \rangle$,
if it is not vanishing, belonging to the $\mu$ row of the 
representation $\Gamma$ of the point group $G$:
$$R\psi^{\Gamma}_{\mu \nu}~|j, \rho \rangle=\displaystyle \sum_{\lambda}~
D^{\Gamma}_{\lambda \mu}(R)\psi^{\Gamma}_{\lambda \nu}|j, \rho \rangle.
\eqno (6) $$

\noindent
This method is effective for both integer and half-odd-integer
angular momentum states. In this paper we will calculate the
irreducible bases in the group space of the octahedral 
double group {\bf O'} (Sec. 2), and then, find a simple and 
unified formula (see (18) in Sec.3) for calculating the SAB. This 
method is effective for all double point groups. Most double 
point groups are the subgroups of {\bf O'}. The SAB for a
subgroup can also be obtained from SAB of {\bf O'} by reducing
the subduced representations of {\bf O'} for the subgroup.
In Sec. 4 we will demonstrate this method by taking the
tetrahedral double group {\bf T'} as an example. The calculation
for the icosahedral double group will be published elsewhere
\cite{dong}.  At last, in Sec. 5 we will give some conclusions.

\vspace{4mm}
\noindent
{\bf 2. Octahedral double group}

\vspace{3mm}
A cube is shown in Fig.1. The vertices on the upper part are 
labeled by $A_{j}$, $1\leq j \leq 4$, and their opposite 
vertices by $B_{j}$. The coordinate axes point from the center
$O$ to the centers of the faces, respectively.

\begin{center}
\fbox{Fig. 1.}
\end{center}

The group {\bf O} contains 3 four-fold axes, 4 three-fold axes, and 6 
two-fold axes. The four-fold axes are along the coordinate axes, 
and the rotations through $\pi/2$ around those four-fold axes are 
denoted by $T_{x}$, $T_{y}$, and $T_{z}$, respectively. 
The three-fold axes  point from $B_{j}$ to $A_{j}$ ($1 \leq j \leq 4$) 
with the polar angle $\theta$ and azimuthal angles $\varphi_{j}$:
$$\cos \theta=\sqrt{1/3},~~~~\varphi_{j}=(2j-1)\pi/4,~~~~1\leq j \leq 4.
\eqno (7) $$

\noindent
The rotations through $2\pi/3$ around those three-fold axes are 
denoted by $R_{j}$, $1\leq j \leq 4$. The two-fold axes join the 
midpoints of two opposite edges, and corresponding rotations are
denoted by $S_{j}$, $1\leq j \leq 6$. The polar and azimuthal angles 
of the first 4 axes are $\pi/4$ and $(j-1)\pi/2$, and the last two 
axes are located on the $xy$ plane with the azimuthal angles $\pi/4$ and
$3\pi/4$, respectively.

The octahedral double group {\bf O'} contains 48 elements and
eight classes. There are eight inequivant irreducible representations
for {\bf O'}: Five representations $D^{A_{1}}$, $D^{A_{2}}$, $D^{E}$, 
$D^{T_{1}}$, and $D^{T_{2}}$ are called single-valued ones, and three 
representations $D^{E_{1}'}$, $D^{E_{2}'}$, and $D^{G'}$ are 
double-valued ones. From a standard
calculation of group theory, the character table is obtained and listed in
Table 1. The row (column) index $\mu$ runs over integer (in a single-valued 
representation) or half-odd-integer (in a double-valued one). The order
of the row index $\mu$ are also listed in Table 1.

\begin{center}
\fbox{Table 1}
\end{center}

The octahedral double group {\bf O$_{h}'$} is the direct product of
{\bf O'} and the inversion group $\{E,P\}$, where $P$ is the 
inversion operator. According to the parity, the irreducible 
representations of {\bf O$_{h}'$} are denoted as $\Gamma_{g}$
(even) and $\Gamma_{u}$ (odd), respectively. In this paper we 
will pay more attention to the double group {\bf O'}. 

The rank of the double group {\bf O'} is three. We choose 
$E'$, $T_{z}$, and $S_{1}$ as the generators of {\bf O'}. 
The representation matrix of $E'$ is equal to the unit
matrix {\bf 1} in a single-valued irreducible representation
and {\bf $-1$} in a double-valued one. It is convenient to 
choose the bases in an irreducible representations
of {\bf O'} such that the representation matrices of the 
generator $T_{z}$ are diagonal with the diagonal elements 
$\eta^{\mu}$, where $\eta=\exp\{-i\pi/2\}$. Assume that the 
bases $\Phi_{\mu \nu}$ in the {\bf O'} group space are the 
eigenstates of left-action and right-action of $T_{z}$:
$$\begin{array}{ll}
T_{z}~\Phi_{\mu \nu}=\eta^{\mu}\Phi_{\mu \nu},~~~~~
&\Phi_{\mu \nu}~T_{z}=\eta^{\nu}\Phi_{\mu \nu}, \\
\end{array} \eqno (8) $$

\noindent
The bases $\Phi_{\mu \nu}$ can be easily calculated by the 
projection operator $P_{\mu}$ (see p.113 in \cite{ham}): 
$$\Phi_{\mu \nu}=c~P_{\mu}~R~P_{\nu},~~~~~
P_{\mu}=\displaystyle {E+\eta^{-4\mu}E' \over 8} \sum_{a=0}^{3}~
\eta^{-\mu a}~T_{z}^{a}, \eqno (9) $$

\noindent
where $c$ is a normalization factor. The choice of the
group element $R$ in (9) will not affect the results except for
the factor $c$. The subscripts $\mu$ and $\nu$ should be
integer or half-odd-integer, simultaneously. In the following 
we choose $E$, $T_{x}^{2}$ and $S_{1}$ as the group 
element $R$ in (9), respectively, and obtain three independent sets 
of bases $\Phi_{\mu \nu}^{(i)}$: 
$$\begin{array}{rl}
\Phi^{(1)}_{\mu \mu}&=\displaystyle {E+\eta^{-4\mu}E' \over 2\sqrt{2}} 
\sum_{a=0}^{3}~\eta^{-\mu a}~T_{z}^{a}\\
\Phi^{(2)}_{\mu \overline{\mu}}&=\displaystyle 
{E+\eta^{-4\mu}E' \over 2\sqrt{2}} 
\sum_{a=0}^{3}~\eta^{-\mu a}~T_{z}^{a}T_{x}^{2}\\
&=\displaystyle {E+\eta^{-4\mu}E' \over 2\sqrt{2}} 
\left(T_{x}^{2}+\eta^{-\mu}S_{5}+\eta^{-2\mu} T_{y}^{2} 
+\eta^{-3\mu}S_{6}\right), \\
\end{array} $$
$$\begin{array}{rl}
\Phi^{(3)}_{\mu \nu}&=\displaystyle {E+\eta^{-4\mu}E' \over 4\sqrt{2}} 
\sum_{a=0}^{3}~\eta^{-\mu a}~T_{z}^{a}S_{1} \sum_{b=0}^{3}~
\eta^{-\nu b}~T_{z}^{b}\\
&=\displaystyle {E+\eta^{-4\mu}E' \over 4\sqrt{2}}
\left\{\left(S_{1}+\eta^{-\mu}R^{2}_{1}
+\eta^{-2 \mu}T^{3}_{y}+\eta^{\mu}R_{4} \right)\right.\\
&~~~+~\eta^{(\mu-\nu)}\left(S_{4}+\eta^{-\mu}R^{2}_{4}
+\eta^{-2 \mu}T^{3}_{x}+\eta^{\mu}R_{3} \right)\\
&~~~+~\eta^{2(\mu-\nu)}\left(S_{3}+\eta^{-\mu}R^{2}_{3}
+\eta^{2 \mu}T_{y}+\eta^{\mu} R_{2} \right)\\
&\left.~~~+~\eta^{3(\mu-\nu)}\left(S_{2}+\eta^{-\mu}R^{2}_{2}
+\eta^{2 \mu}T_{x}+\eta^{\mu} R_{1} \right)\right\}. 
 \end{array} \eqno (10) $$

\noindent
where and hereafter the subscript $\overline{\mu}$ denotes $-\mu$. 
Those bases $\Phi_{\mu \nu}^{(i)}$ should
be combined into the irreducible bases $\psi_{\mu \nu}^{\Gamma}$ 
belonging to the given irreducible representation $\Gamma$. The 
combinations can be determined from the condition that the 
irreducible bases should be the eigenstate of a class operator $W$, 
which was called CSCO-III in \cite{chen}. The eigenvalues 
$\alpha_{\Gamma}$ can be calculated (see (3-170) in \cite{ham}) 
from the characters in the irreducible representations $\Gamma$ 
listed in Table 1:
$$\begin{array}{ll}
W=T_{x}+T_{y}+T_{z}+E'T_{x}^{3}+E'T_{y}^{3}+E'T_{z}^{3},~~
&W~\psi_{\mu \nu}^{\Gamma}=\psi_{\mu \nu}^{\Gamma}~W=
\alpha_{\Gamma}~\psi_{\mu \nu}^{\Gamma}, \\
\alpha_{A_{1}}=6,~~~~~~~~~~~~\alpha_{A_{2}}=-6,
&\alpha_{E}=0,~~~~~~~~~~\alpha_{T_{1}}=2, \\
\alpha_{T_{2}}=-2,~~~~~~~~~~\alpha_{E_{1}'}=3\sqrt{2},
&\alpha_{E_{2}'}=-3\sqrt{2},~~~~\alpha_{G'}=0.
\end{array} \eqno (11) $$

\noindent
Although a coincidence occurs, $\alpha_{E}=\alpha_{G'}$,
this coincidence will not constitute an obstacle against
calculation, because $D^{E}$ is a single-valued representation, 
but $D^{G'}$ is a double-valued one. 

Now we are able to calculate the matrix form of $W$ in the bases 
$\Phi_{\mu \nu}^{(i)}$, and diagonalize it. $\psi_{\mu \nu}^{\Gamma}$
are just the eigenvectors of the matrix form of $W$:
$$\psi_{\mu \nu}^{\Gamma}=N^{-1/2}~\displaystyle \sum_{i=1}^{3}~
C_{i}~\Phi_{\mu \nu}^{(i)} , \eqno (12) $$

\noindent
where $N$ is the normalization factor. 

In these irreducible bases the representation matrices of 
$E'$ and $T_{z}$ are diagonal with the diagonal elements 
$\pm 1$ and $\eta^{\mu}$, respectively. But the explicit
matrix forms of another generator $S_{1}$ will depend 
upon the phases of the bases $\psi_{\mu \nu}^{\Gamma}$.
We choose the phases such that $S_{1}$ has the following
representation matrices:
$$\begin{array}{ll}
D^{A_{1}}(S_{1})=-D^{A_{2}}(S_{1})=1,~~~~
&D^{E}(S_{1})=\displaystyle {1 \over 2}\left(\begin{array}{cc} 
1 & \sqrt{3} \\ \sqrt{3} & -1 \end{array} \right),\\
\end{array} $$
$$\begin{array}{l}
D^{T_{1}}(S_{1})=-D^{T_{2}}(S_{1})
=\displaystyle {1 \over 2}\left(\begin{array}{ccc} 
-1 & -\sqrt{2} & -1 \\ -\sqrt{2} & 0 & \sqrt{2} \\
-1 & \sqrt{2} & -1 \end{array} \right), \\
D^{E_{1}'}(S_{1})=\displaystyle {i \over \sqrt{2}}\left(\begin{array}{cc} 
-1 & -1 \\ -1 & 1 \end{array} \right),~~~~~~
D^{E_{2}'}(S_{1})=\displaystyle {i \over \sqrt{2}}\left(\begin{array}{cc} 
-1 & 1 \\ 1 & 1 \end{array} \right),\\
D^{G'}(S_{1})=\displaystyle {i \over 2\sqrt{2}}\left(\begin{array}{cccc} 
1 & \sqrt{3} & \sqrt{3} & 1 \\ \sqrt{3} & 1 & -1 & -\sqrt{3} \\
\sqrt{3} & -1 & -1 & \sqrt{3} \\ 1 & -\sqrt{3} & \sqrt{3} & -1 
 \end{array} \right).
\end{array}  \eqno (13) $$

\noindent
Some representations coincide with the subduced representations 
of $D^{j}$ of SO(3):
$$\begin{array}{ll}
D^{A_{1}}(R)=D^{0}(R),~~~~&D^{T_{1}}(R)=D^{1}(R), \\
D^{E_{1}'}(R)=D^{1/2}(R),~~&D^{G'}(R)=D^{3/2}(R),  
\end{array} ~~~~~~R\in {\bf O'}. \eqno (14) $$

The normalization factors $N$ and combination
coefficients $C_{i}$ in the irreducible bases (12)
are listed in Table 2. 

\begin{center}
\fbox{Table 2}
\end{center}

Now, the irreducible bases $\psi_{\mu \nu}^{\Gamma}$ satisfy
(5). The irreducible bases of the group {\bf O$_{h}'$} 
can be expressed as follows:
$$\psi_{\mu \nu}^{\Gamma_{g}}=2^{-1/2} \left(E+P\right)
\psi_{\mu \nu}^{\Gamma},~~~~~~~
\psi_{\mu \nu}^{\Gamma_{u}}=2^{-1/2} \left(E-P\right)
\psi_{\mu \nu}^{\Gamma}. \eqno (15) $$

\vspace{4mm}
\noindent
{\bf 3. Applications to the angular momentum states}

\vspace{3mm}
Due to the properties (5), we can obtain the SAB by applying 
$\psi_{\mu \nu}^{\Gamma}$ to any function. As an important 
application, we apply $\psi_{\mu \nu}^{\Gamma}$ to the angular 
momentum states $|j,\mu \rangle$, where the Condon-Shortley definition
is used:
$$R~|j,\mu \rangle=\displaystyle \sum_{\nu=-j}^{j}~
D^{j}_{\nu \mu}(R)~|j, \nu \rangle,~~~~R\in SO(3) {\rm ~or~} SU(2).
\eqno (16) $$ 

\noindent
When $j$ is an integer $\ell$, $|\ell,m\rangle$ is just the 
spherical harmonics $Y^{\ell}_{m}(\theta, \varphi)$. 

From (16) and the definitions of the group elements we have: 
$$\begin{array}{l}
E'~|j, \mu \rangle=(-1)^{2j}~|j, \mu \rangle,~~~~~~
T_{z}~|j, \mu \rangle=\eta^{\mu}~|j, \mu \rangle,\\
T_{x}^{2}~|j,\mu \rangle =\displaystyle \sum_{\nu}~
D^{j}_{\nu \mu}(0, \pi, \pi)~|j,\nu \rangle
=(-1)^{j-\mu}\eta^{2\mu}~|j,-\mu \rangle , \\
S_{1}~|j,\mu \rangle =\displaystyle \sum_{\nu}~
D^{j}_{\nu \mu}(0, \pi/2, \pi)~|j,\nu \rangle
=\displaystyle \sum_{\nu}~
\eta^{2\mu}d^{j}_{\nu \mu}(\pi/2)~|j,\nu \rangle .
\end{array} \eqno (17) $$

Now, it is easy to obtain the combinations of 
the angular momentum states
$\psi_{\mu \lambda}^{\Gamma}~|j,\rho \rangle$ into SAB 
of {\bf O'}:
$$\psi_{\mu \lambda}^{\Gamma}~|j,\rho \rangle =\sqrt{8/N}
\delta_{\lambda \rho}' \displaystyle \sum_{\nu}~\delta_{\mu \nu}'
~|j,\nu\rangle \left\{C_{1}\delta_{\rho \nu}+C_{2}
\delta_{\overline{\rho} \nu}(-1)^{j-\rho} \eta^{2\rho}
+2C_{3}\eta^{2\rho}d^{j}_{\nu \rho}(\pi/2) \right\} .
 \eqno (18) $$

\noindent
where $N$ and $C_{i}$ were given in Table 2, $\eta=\exp\{-i\pi/2\}$, 
and $\delta_{\lambda \rho}'$ is defined as follows:
$$\delta_{\lambda \rho}'=\left\{\begin{array}{ll}1~~~~&{\rm when}~~
(\lambda-\rho)/n={\rm integer} \\
0 & {\rm otherwise}. \end{array} \right. \eqno (19) $$

\noindent
where $n=4$ for {\bf O'} due to $\eta^{4}=1$.
In driving (18) some terms were merged so that the functions
need be normalized again.

(18) is a simple and unified formula for calculating the
correlations of the spin states. For the fixed $\lambda$ 
and $\rho$, satisfying $\delta_{\lambda \rho}'=1$,
we obtain the combinations of the angular momentum states
$\psi_{\mu \lambda}^{\Gamma}~|j,\rho \rangle$, belonging to the $\mu$
row of the irreducible representation $\Gamma$ of {\bf O'}. 
Different choice of $\lambda$ and $\rho$ may cause the combinations
vanishing, dependent on each other, or independent. The
number of independent combinations depends upon the number of times
that the irreducible representation $\Gamma$ of {\bf O'} is 
contained in the subduced representation of $D^{j}$ of SU(2). 
The latter is completely determined by the 
characters of the representations $\Gamma$ and $D^{j}$.

Those combinations given in (18) are very easy to be calculated, 
by a simple computer file or even by hand. In the following we list
some combinations as examples.
$$\psi_{00}^{A}~|0,0\rangle =4\sqrt{3}~|0,0\rangle,~~~~~
\psi_{\mu 1}^{T_{1}}~|1, 1\rangle =4~|1,\mu\rangle. $$
$$\begin{array}{l}
\psi_{2 2}^{E}~|2, 2\rangle =2\sqrt{3}~\left(\sqrt{1/2}~|2,2 \rangle
+\sqrt{1/2}~|2,-2 \rangle \right), \\
\psi_{0 2}^{E}~|2, 2\rangle =2\sqrt{3}~|2, 0 \rangle, \\
\psi_{3 2}^{T_{2}}~|2, 2\rangle =2\sqrt{2}~|2,-1 \rangle, \\
\end{array} $$
$$\begin{array}{l}
\psi_{2 2}^{T_{2}}~|2, 2\rangle =2\sqrt{2}~\left(\sqrt{1/2}~|2,2 \rangle
-\sqrt{1/2}~|2,-2 \rangle \right), \\
\psi_{1 2}^{T_{2}}~|2, 2\rangle =2\sqrt{2}~\left(-|2,1 \rangle \right). \\
\psi_{2 2}^{A_{2}}~|3, 2\rangle =2\sqrt{6}~\left(\sqrt{1/2}~|3,2 \rangle
-\sqrt{1/2}~|3,-2 \rangle \right), \\
\psi_{1 \overline{1}}^{T_{1}}~|3, 3\rangle =\sqrt{10}~\left(
\sqrt{3/8}~|3,1 \rangle +\sqrt{5/8}~|3,-3 \rangle \right), \\
\psi_{0 \overline{1}}^{T_{1}}~|3, 3\rangle =\sqrt{10}~\left(-|3,0 \rangle
\right), \\
\psi_{\overline{1}\, \overline{1}}^{T_{1}}~|3, 3\rangle =\sqrt{10}~
\left(\sqrt{5/8}~|3,3 \rangle +\sqrt{3/8}~|3,-1 \rangle \right), \\
\psi_{3 \overline{1}}^{T_{2}}~|3, 3\rangle =-\sqrt{6}~\left(
-\sqrt{3/8}~|3,3 \rangle +\sqrt{5/8}~|3,-1 \rangle \right), \\
\psi_{2 \overline{1}}^{T_{2}}~|3, 3\rangle =-\sqrt{6}~\left(
\sqrt{1/2}~|3,2 \rangle +\sqrt{1/2}~|3,-2 \rangle \right), \\
\psi_{1 \overline{1}}^{T_{2}}~|3, 3\rangle =-\sqrt{6}~\left(
\sqrt{5/8}~|3,1 \rangle -\sqrt{3/8}~|3,-3 \rangle \right). \\
\end{array} $$
$$\psi_{\mu (1/2)}^{E_{1}'}~|1/2, 1/2\rangle =i2\sqrt{6}~|1/2,\mu\rangle, ~~~~
\psi_{\mu (3/2)}^{G'}~|3/2, 3/2\rangle =i2\sqrt{3}~|3/2,\mu\rangle. $$
$$\begin{array}{l}
\psi_{(3/2)\,(\overline{3/2})}^{E_{2}'}~|5/2, 5/2\rangle =i2~\left(
-\sqrt{5/6}~|5/2,3/2 \rangle +\sqrt{1/6}~|5/2, -5/2 \rangle \right), \\
\psi_{(\overline{3/2})\,(\overline{3/2})}^{E_{2}'}~|5/2, 5/2\rangle =i2~\left(
\sqrt{1/6}~|5/2,5/2 \rangle -\sqrt{5/6}~|5/2, -3/2 \rangle \right), \\
\psi_{(3/2)\,(\overline{3/2})}^{G'}~|5/2, 5/2\rangle =i\sqrt{10}~\left(
-\sqrt{1/6}~|5/2,3/2 \rangle -\sqrt{5/6}~|5/2, -5/2 \rangle \right), \\
\psi_{(1/2)\,(\overline{3/2})}^{G'}~|5/2, 5/2\rangle =i\sqrt{10}~
|5/2,1/2 \rangle, \\
\psi_{(\overline{1/2})\,(\overline{3/2})}^{G'}~|5/2, 5/2\rangle 
=i\sqrt{10}~\left(-|5/2,-1/2 \rangle \right) , \\
\psi_{(\overline{3/2})\,(\overline{3/2})}^{G'}~|5/2, 5/2\rangle 
=i\sqrt{10}~\left(
\sqrt{5/6}~|5/2,5/2 \rangle +\sqrt{1/6}~|5/2, -3/2 \rangle \right). \\
\end{array} $$

\vspace{4mm}
\noindent
{\bf 4. Tetrahedral double group}

\vspace{3mm}
The tetrahedral double group {\bf T'} is a subgroup of
{\bf O'} with the generators $E'$, $T_{z}^{2}$, and $R_{1}$:
$$R_{1}=E'S_{1}T_{z}^{3}=S_{1}T_{z}^{-1}. \eqno (20) $$

\noindent
In the irreducible bases we choose, the representation matrices of 
$E'$ and $T_{z}^{2}$ are diagonal with the diagonal elements 
$\pm 1$ and $\eta^{2\mu}$, respectively, and the 
representation matrices of $R_{1}$ are as follows:
$$\begin{array}{l}
~D^{A_{0}}(R_{1})=1,~~~D^{A_{+}}(R_{1})=\omega=\exp\{-i2\pi/3\},~~~
D^{A_{-}}(R_{1})=\overline{\omega}=\exp\{i2\pi/3\}, \\
\begin{array}{ll}
D^{T}(R_{1})=\displaystyle {1 \over 2} \left(\begin{array}{ccc}
-i & -\sqrt{2} & i \\ -i\sqrt{2} & 0 & -i\sqrt{2} \\
-i & \sqrt{2} & i \end{array} \right), ~~
&D^{E_{0}'}(R_{1})=\displaystyle {\tau \over \sqrt{2}} 
\left(\begin{array}{cc} 1 & -i \\ 1 & i \end{array} \right),\\
D^{E_{+}'}(R_{1})=\omega D^{E_{0}'}(R_{1}),~~~~
&D^{E_{-}'}(R_{1})=\overline{\omega} D^{E_{0}'}(R_{1}). 
\end{array} \end{array}  \eqno (21) $$

\noindent
where $\tau=\exp\{-i\pi/4\}$. 
The character table of {\bf T'} is given in Table 3. 

\begin{center}
\fbox{Table 3}
\end{center}
 
Generally speaking, an irreducible representation of {\bf O'} 
is a reducible one as a subduced representation of {\bf T'}.
Through the following similarity transformation, the SAB 
of {\bf O'} can be further combined into the SAB of {\bf T'}:
$$\begin{array}{l}
D^{A_{1}}(R),~~D^{A_{2}}(R)~\longrightarrow~D^{A_{0}}(R), \\
X_{1}^{-1}D^{E}(R)X_{1}~\longrightarrow~D^{A_{+}}(R) \oplus D^{A_{-}}(R), \\
D^{T_{1}}(R),~~D^{T_{2}}(R)~\longrightarrow~D^{T}(R), \\
D^{E_{1}'}(R),~~X_{2}^{-1}D^{E_{2}'}(R)X_{2}
~\longrightarrow~D^{E_{0}'}(R), \\
X_{3}^{-1}D^{G'}(R)X_{3}~
\longrightarrow~D^{E_{+}'}(R) \oplus D^{E_{-}'}(R), \\
X_{1}=\displaystyle { 1 \over \sqrt{2} }
\left( \begin{array}{cc} 1 & 1 \\ -i & i \end{array} \right),~~~~
X_{2}=\left( \begin{array}{cc} 0 & 1 \\ 1 & 0 \end{array} \right), \\
X_{3}=\displaystyle { 1 \over \sqrt{2} }
\left( \begin{array}{cccc} 0 & 1 & 0 & 1 \\ i & 0 & -i & 0 \\
0 & -i & 0 & i \\ -1 & 0 & -1 & 0 \end{array} \right).
\end{array}  \eqno (22) $$

\noindent
For example, 
$$\begin{array}{rl}
\psi_{00}^{'A_{+}}~|2,2\rangle &=~\left\{\psi_{22}^{E}~|2,2\rangle
-i\psi_{02}^{E}~|2,2\rangle\right\}/\sqrt{2} \\
&\sim~\left\{|2,2\rangle -i \sqrt{2} |2,0 \rangle + |2,-2 \rangle
\right\}/2, \\
\psi_{00}^{'A_{-}}~|2,2\rangle &=~\left\{ \psi_{22}^{E}~|2,2\rangle
+i \psi_{02}^{E}~|2,2\rangle\right\}/\sqrt{2} \\
&\sim~\left\{|2,2\rangle + i\sqrt{2} |2,0 \rangle + |2,-2 \rangle
\right\}/2, \\
\psi_{10}^{'T}~|2,2\rangle &= ~\psi_{32}^{T_{2}}~|2,2\rangle
\sim |2,-1 \rangle, \\
\psi_{00}^{'T}~|2,2\rangle &=~\psi_{22}^{T_{2}}~|2,2\rangle
\sim\left\{|2,2\rangle - |2,-2 \rangle
\right\}/\sqrt{2}, \\
\psi_{\overline{1}0}^{'T}~|2,2\rangle  &= ~\psi_{12}^{T_{2}}~|2,2\rangle
\sim -|2,1 \rangle.
\end{array} \eqno (23) $$

As an alternative method, (23) can also be obtained from
the irreducible bases of the group space of {\bf T'}. 
Similar to (9) and (10), we calculate the bases $\Phi_{\mu \nu}^{'(i)}$
by the projection operator $P_{\mu}'$:
$$\Phi_{\mu \nu}'=c~P_{\mu}'~R~P_{\nu}',~~~~~ 
P_{\mu}'=\displaystyle {1 \over 4} ~\left(E+\eta^{-4\mu}E'\right)
\left(E+\eta^{-2\mu}T_{z}^{2}\right). \eqno (24) $$

\noindent
We choose $E$, $T_{x}^{2}$, $R_{1}$, and $R_{1}^{2}$ as the group
element $R$ in (24), respectively, and obtain four
independent sets of bases $\Phi_{\mu \nu}^{'(i)}$:
$$\begin{array}{l}
\Phi_{\mu \mu}^{'(1)}=\displaystyle {E+\eta^{4\mu}E' \over 2}~
\left(E+\eta^{-2\mu} T_{z}^{2} \right), \\
\Phi_{\mu \overline{\mu}}^{'(2)}=\displaystyle {E+\eta^{4\mu}E' \over 2}~
\left(T_{x}^{2}+\eta^{-2\mu} T_{y}^{2} \right), \\
\Phi_{\mu \nu}^{'(3)}=\displaystyle {E+\eta^{4\mu}E' \over 2\sqrt{2}}~
\left(R_{1}+\eta^{-2\mu} R_{2}^{2}
+\eta^{2(\mu-\nu)}R_{3}+\eta^{-2\nu} R_{4}^{2} \right), \\
\Phi_{\mu \nu}^{'(4)}=\displaystyle {E+\eta^{4\mu}E' \over 2\sqrt{2}}~
\left(R_{1}^{2}+\eta^{2\mu} R_{4}
+\eta^{2(\mu-\nu)}R_{3}^{2}+\eta^{2\nu} R_{2} \right).
\end{array} \eqno (25) $$

\noindent
The class operator $W'$ is used to determine the irreducible
bases $\psi_{\mu \nu}^{'\Gamma}$:
$$\begin{array}{ll}
W'=R_{1}+R_{3}+E'R_{1}^{2}+E'R_{3}^{2},~~
&W'~\psi_{\mu \nu}^{'\Gamma}=\psi_{\mu \nu}^{'\Gamma}~W'=
\beta_{\Gamma}~\psi_{\mu \nu}^{'\Gamma}, \\
\beta_{A_{0}}=4,~~~~~~~~\beta_{A_{+}}=4\omega,~~~~~~
&\beta_{A_{-}}=4\overline{\omega},~~~~\beta_{T}=0, \\
\beta_{E_{0}'}=2,~~~~~~~~\beta_{E_{+}'}=2\omega,~~~~~~
&\beta_{E_{-}'}=2\overline{\omega}.
\end{array} \eqno (26) $$

\noindent
$\psi_{\mu \nu}^{'\Gamma}$ are the eigenvectors of the matrix 
form of $W'$ in the bases $\Phi_{\mu \nu}^{'(i)}$:
$$\psi_{\mu \nu}^{'\Gamma}=N^{-1/2} \displaystyle 
\sum_{i=1}^{4}~C_{i}~\Phi_{\mu \nu}^{'\Gamma}, \eqno (27) $$

\noindent
where the normalization factor $N$ and the combination coefficients 
$C_{i}$ are listed in Table 4.

\begin{center}
\fbox{Table 4}
\end{center}

Recall the Eulerian angles of the relevant rotations:
$$T_{x}^{2}=R(0, \pi, \pi),~~~~R_{1}=R(0, \pi/2, \pi/2),~~~~
R_{1}^{2}=R(\pi/2, \pi/2, \pi). \eqno (28) $$

\noindent
Now, applying the irreducible bases $\psi_{\mu \nu}^{'\Gamma}$
to the angular momentum states $|j, \mu \rangle$, we
obtain the SAB for {\bf T'}:
$$\begin{array}{rl}
\psi_{\mu \lambda}^{'\Gamma}~|j,\rho \rangle &=~\sqrt{4/N}
\delta_{\lambda \rho}' \displaystyle \sum_{\nu}~\delta_{\mu \nu}'
~|j, \nu \rangle \left\{C_{1}\delta_{\rho \nu}
+C_{2}\delta_{\overline{\rho} \nu}(-1)^{j-\rho} \eta^{2\rho}\right. \\
&\left.~~~+\sqrt{2}C_{3}\eta^{\rho}d^{j}_{\nu \rho}(\pi/2) 
+\sqrt{2}C_{4}\eta^{\nu+2\rho}d^{j}_{\nu \rho}(\pi/2) \right\}.
\end{array} \eqno (29) $$

\noindent
where $\delta_{\lambda \rho}'$ is defined in (19) with $n=2$. 

It is very easy to calculate from (29) the combinations of 
the angular momentum states $|j, \mu \rangle$ into the SAB 
of {\bf T'}. In the following we list some examples:
$$\psi_{00}^{'A_{0}}~|0,0\rangle =2\sqrt{6}~|0,0\rangle,~~~~~
\psi_{\mu 1}^{'T}~|1, 1\rangle =2\sqrt{2}~|1,\mu\rangle. $$
$$\begin{array}{l}
\psi_{00}^{'A_{+}}~|2, 2\rangle =\sqrt{6}~\left((1/2)~|2,2 \rangle
-i\sqrt{1/2}~|2,0\rangle +(1/2)~|2,-2 \rangle \right), \\
\psi_{00}^{'A_{-}}~|2, 2\rangle =\sqrt{6}~\left((1/2)~|2,2 \rangle
+i\sqrt{1/2}~|2,0\rangle +(1/2)~|2,-2 \rangle \right), \\
\psi_{10}^{'T}~|2, 2\rangle =2~|2,-1 \rangle, \\
\psi_{00}^{'T}~|2, 2\rangle =2~\left(\sqrt{1/2}~|2,2 \rangle
-\sqrt{1/2}~|2,-2 \rangle \right), \\
\psi_{\overline{1} 0}^{T}~|2, 2\rangle =2~\left(-|2,1 \rangle \right). \\
\end{array} $$
$$\begin{array}{l}
\psi_{00}^{A_{0}}~|3, 2\rangle =2\sqrt{3}~\left(\sqrt{1/2}~|3,2 \rangle
-\sqrt{1/2}~|3,-2 \rangle \right), \\
\psi_{11}^{T}~|3, 3\rangle =-\sqrt{3}~\left(
-\sqrt{3/8}~|3,3 \rangle +\sqrt{5/8}~|3,-1 \rangle \right), \\
\psi_{01}^{T}~|3, 3\rangle =-\sqrt{3}~\left(\sqrt{1/2}~|3,2 \rangle
+\sqrt{1/2}~|3,-2 \rangle \right), \\
\psi_{\overline{1} 1}^{T}~|3, 3\rangle =-\sqrt{3}~
\left(\sqrt{5/8}~|3,1 \rangle -\sqrt{3/8}~|3,-3 \rangle \right), \\
\psi_{1 \overline{1}}^{T}~|3, 3\rangle =\sqrt{5}~\left(
\sqrt{3/8}~|3,1 \rangle +\sqrt{5/8}~|3,-3 \rangle \right), \\
\psi_{0 \overline{1}}^{T}~|3, 3\rangle =\sqrt{5}~\left(-|3,0 \rangle
\right) , \\
\psi_{\overline{1}\, \overline{1}}^{T}~|3, 3\rangle =\sqrt{5}~\left(
\sqrt{5/8}~|3,3 \rangle +\sqrt{3/8}~|3,-1 \rangle \right). \\
\end{array} $$
$$\begin{array}{l}
\psi_{\mu (1/2)}^{E_{0}'}~|1/2, 1/2\rangle =2\sqrt{3}\tau~|1/2,\mu\rangle,\\
\psi_{(1/2)(\overline{1/2})}^{E_{+}'}~|3/2, 3/2\rangle 
=\sqrt{6}\tau~\left(i\sqrt{1/2}~|3/2,1/2\rangle-\sqrt{1/2}~|3/2,-3/2
\rangle \right) , \\
\psi_{(\overline{1/2})\,(\overline{1/2})}^{E_{+}'}~|3/2, 3/2\rangle 
=\sqrt{6}\tau~\left(\sqrt{1/2}~|3/2,3/2\rangle-i\sqrt{1/2}~|3/2,-1/2
\rangle \right), \\
\psi_{(1/2)(\overline{1/2})}^{E_{-}'}~|3/2, 3/2\rangle 
=\sqrt{6}\tau~\left(-i\sqrt{1/2}~|3/2,1/2\rangle-\sqrt{1/2}~|3/2,-3/2
\rangle \right), \\
\psi_{(\overline{1/2})\,(\overline{1/2})}^{E_{-}'}~|3/2, 3/2\rangle 
=\sqrt{6}\tau~\left(\sqrt{1/2}~|3/2,3/2\rangle+i\sqrt{1/2}~|3/2,-1/2
\rangle \right), \\
\end{array} $$
$$\begin{array}{rl}
\psi_{(1/2)(1/2)}^{E_{0}'}~|5/2, 5/2\rangle 
&=~\sqrt{2}\tau~\left(\sqrt{1/6}~|5/2,5/2\rangle-\sqrt{5/6}~|5/2,-3/2
\rangle \right), \\
\psi_{(\overline{1/2})(1/2)}^{E_{0}'}~|5/2, 5/2\rangle 
&=~\sqrt{2}\tau~\left(-\sqrt{5/6}~|5/2,3/2\rangle+\sqrt{1/6}~|5/2,-5/2
\rangle \right), \\
\psi_{(1/2)(1/2)}^{E_{+}'}~|5/2, 5/2\rangle 
&=~\sqrt{5}\tau~\left(\sqrt{5/12}~|5/2,5/2\rangle-i\sqrt{1/2}~|5/2,1/2
\rangle \right. \\
&\left.~~~+\sqrt{1/12}~|5/2,-3/2 \rangle \right), \\
\psi_{(\overline{1/2})(1/2)}^{E_{+}'}~|5/2, 5/2\rangle 
&=~\sqrt{5}\tau~\left(\sqrt{1/12}~|5/2,3/2\rangle-i\sqrt{1/2}~|5/2,-1/2
\rangle \right.\\
&\left.~~~+\sqrt{5/12}~|5/2,-5/2 \rangle \right), \\
\end{array} $$
$$\begin{array}{rl}
\psi_{(1/2)(1/2)}^{E_{-}'}~|5/2, 5/2\rangle 
&=~\sqrt{5}\tau~\left(\sqrt{5/12}~|5/2,5/2\rangle+i\sqrt{1/2}~|5/2,1/2
\rangle \right. \\
&\left.~~~+\sqrt{1/12}~|5/2,-3/2 \rangle \right), \\
\psi_{(\overline{1/2})(1/2)}^{E_{-}'}~|5/2, 5/2\rangle 
&=~\sqrt{5}\tau~\left(\sqrt{1/12}~|5/2,3/2\rangle+i\sqrt{1/2}~|5/2,-1/2
\rangle\right. \\
&\left.~~~+\sqrt{5/12}~|5/2,-5/2 \rangle \right), \\
\end{array} $$

\noindent
It is obvious that (23) coincides with these combinations.

\vspace{4mm}
\noindent
{\bf 5. Conclusion}

\vspace{3mm}
The eigenstates of the Hamiltonian of a system with
a given symmetry can be combined into the symmetry adapted 
bases \cite{ham}. From the irreducible bases in the group space of 
the symmetry group of the system, the symmetry adapted bases 
can be calculated generally and simply. The combinations of 
the angular momentum states are important examples for 
calculating the symmetry adapted bases. In this paper we 
calculate the explicit form of the irreducible bases of {\bf O'} 
group space, and obtain a general formula (18) for calculating the 
combinations of angular momentum states into the SAB of {\bf O'}.
These method is effective for all double point groups. 
However, most double point groups are the subgroups of 
{\bf O'}, and the SAB of a subgroup of {\bf O'} can also
be calculated by further combining the SAB of {\bf O'}.
The calculation for the icosahedral double group will be 
published elsewhere \cite{dong}.  

\vspace{5mm}
{\bf Acknowledgments}. The authors would like to thank professor 
Jin-Quan Chen for sending us his preprint \cite{chen} before 
publication. This work was supported by 
the National Natural Science Foundation of China and Grant No. 
LWTZ-1298 of Chinese Academy of Sciences.

\vspace{20mm}
\begin{center}
\setlength{\unitlength}{0.8mm}
\begin{picture}(50,82)(0,-7)

\put(42,49){\makebox(0,0)[l]{\scriptsize $A_{1}$}}
\put(50,72){\makebox(0,0)[b]{\scriptsize $A_{2}$}}
\put(10,72){\makebox(0,0)[b]{\scriptsize $A_{3}$}}
\put(-2,50){\makebox(0,0)[r]{\scriptsize $A_{4}$}}
\put(40,8){\makebox(0,0)[t]{\scriptsize $B_{3}$}}
\put(52,30){\makebox(0,0)[l]{\scriptsize $B_{4}$}}
\put(13,32){\makebox(0,0)[b]{\scriptsize $B_{1}$}}
\put(0,8){\makebox(0,0)[t]{\scriptsize $B_{2}$}}
\put(0,10){\line(1,0){40}}
\put(0,10){\line(0,1){40}}
\put(40,50){\line(-1,0){40}}
\put(40,50){\line(0,-1){40}}
\put(40,50){\line(1,2){10}}
\put(10,70){\line(1,0){40}}
\put(10,70){\line(-1,-2){10}}
\put(50,30){\line(-1,-2){10}}
\put(50,30){\line(0,1){40}}
\put(10,30){\dashbox{1.0}(0,40){}}
\put(10,30){\dashbox{1.0}(40,0){}}
\put(1,12){\circle*{0.5}}
\put(2,14){\circle*{0.5}}
\put(3,16){\circle*{0.5}}
\put(4,18){\circle*{0.5}}
\put(5,20){\circle*{0.5}}
\put(6,22){\circle*{0.5}}
\put(7,24){\circle*{0.5}}
\put(8,26){\circle*{0.5}}
\put(9,28){\circle*{0.5}}

\put(23,40){\makebox(0,0)[r]{\small $O$}}
\put(9.5,5){\makebox(0,0)[l]{$x$}}
\put(55,44){\makebox(0,0)[b]{$y$}}
\put(27,75){\makebox(0,0)[l]{$z$}}
\put(25,40){\vector(1,0){30}}
\put(25,40){\vector(0,1){35}}
\put(25,40){\vector(-1,-2){17.5}}
\put(25,40){\circle*{3}}
\put(20,30){\circle*{3}}
\put(45,40){\circle*{3}}
\put(25,60){\circle*{3}}

\put(25,-5){\makebox(0,0){Fig. 1 $~~$A cube with {\bf O$_{h}$}
symmetry}}

\thicklines
\put(0,10){\line(1,1){40}}
\put(0,10){\line(1,6){10}}
\put(0,10){\line(5,2){50}}
\put(40,50){\line(1,-2){10}}
\put(40,50){\line(-3,2){30}}
\put(10,70){\line(1,-1){40}}

\end{picture}
\end{center}

\vspace{6mm}
\begin{center}

{\bf Table 1} $~~$ Character table of the octahedral double group {\bf O'}

\vspace{3mm}
{\scriptsize
\begin{tabular}{c|cccccccc|c} \hline \hline 
& $E$ & $6C_{4}$ & $6C_{4}^{2}$ & $8C_{3}$ & $12C_{2}$ & $E'$ 
& $6C_{4}^{3}$ & $8C_{3}^{2}$ & $\mu$  \\  \hline
$A_{1}$ & $1$ & $1$ & $1$ & $1$ & $1$ & $1$ & $1$ & $1$ & $0$ \\
$A_{2}$ & $1$ & $-1$ & $1$ & $1$ & $-1$ & $1$ & $-1$ & $1$ & $2$ \\
$E$ & $2$ & $0$ & $2$ & $-1$ & $0$ & $2$ & $0$ & $-1$ & $2,~0$ \\
$T_{1}$ & $3$ & $1$ & $-1$ & $0$ & $-1$ & $3$ & $1$ & $0$ & $1,~0,~-1$ \\
$T_{2}$ & $3$ & $-1$ & $-1$ & $0$ & $1$ & $3$ & $-1$ & $0$ & $3,~2,~1$ \\
$E_{1}'$ & $2$ & $\sqrt{2}$ & $0$ & $1$ & $0$ & $-2$ & $-\sqrt{2}$ 
& $-1$ & $1/2,~-1/2$ \\
$E_{2}'$ & $2$ & $-\sqrt{2}$ & $0$ & $1$ & $0$ & $-2$ & $\sqrt{2}$ 
& $-1$ & $3/2,~-3/2$ \\
$G'$ & $4$ & $0$ & $0$ & $-1$ & $0$ & $-4$ & $0$ &
$1$ & $3/2,~1/2,~-1/2,~-3/2$ \\ \hline \hline
\end{tabular} 
}

\end{center}

\newpage
\begin{center}

{\bf Table 2} $~~$ Irreducible bases in the group space of {\bf O'}
$$\begin{array}{c}
\psi_{\mu \nu}^{\Gamma}=N^{-1/2} \left\{C_{1}~\Phi_{\mu \nu}^{(1)}
+C_{2}~\Phi_{\mu \nu}^{(2)}+C_{3}~\Phi_{\mu \nu}^{(3)} \right\}.
\end{array}  $$
$$\begin{array}{l}
\psi_{00}^{A_{1}}=\left\{\Phi_{00}^{(1)}+\Phi_{00}^{(2)}
+2\Phi_{00}^{(3)} \right\}/\sqrt{6}, \\
\psi_{22}^{A_{2}}=\left\{\Phi_{22}^{(1)}+\Phi_{22}^{(2)}
-2\Phi_{22}^{(3)} \right\}/\sqrt{6}. \\
\end{array} $$

\vspace{2mm}
{\scriptsize
\begin{tabular}{c|c|cccc||c|c|cccc} \hline \hline 
\multicolumn{12}{c}{$\Gamma=E$} \\ \hline
$\mu$ & $\nu$ & $C_{1}$ & $C_{2}$ & $C_{3}$ & $N$ &
$\mu$ & $\nu$ & $C_{1}$ & $C_{2}$ & $C_{3}$ & $N$ \\  \hline
$2$ & $2$ & $1$ & $1$ & $1$ & $3$ &
$2$ & $0$ & & & $1$ & $1$ \\ 
$0$ & $2$ &  &  & $1$ & $1$ &
$0$ & $0$ & $1$ & $1$ & $-1$ & $3$ \\ \hline
\multicolumn{12}{c}{} \\ \hline
\multicolumn{6}{c||}{$\Gamma=T_{1}$} &
\multicolumn{6}{c}{$\Gamma=T_{2}$} \\ \hline
$\mu$ & $\nu$ & $C_{1}$ & $C_{2}$ & $C_{3}$ & $N$ &
$\mu$ & $\nu$ & $C_{1}$ & $C_{2}$ & $C_{3}$ & $N$ \\ \hline
$1$ & $1$ & $1$ & & $-1$ & $2$ &
$3$ & $3$ & $1$ &  & $1$ & $2$ \\ 
$0$ & $1$ &  &  & $-1$ & $1$ &
$2$ & $3$ & & & $1$ & $1$ \\ 
$\overline{1}$ & $1$ & & $-1$ & $-1$ & $2$ &
$1$ & $3$ & & $-1$ & $1$ & $2$ \\
$1$ & $0$ & & & $-1$ & $1$ &
$3$ & $2$ & & & $1$ & $1$ \\
$0$ & $0$ & $1$ & $-1$ & & $2$ &
$2$ & $2$ & $1$ & $-1$ & & $2$ \\
$\overline{1}$ & $0$ & & & $1$ & $1$ &
$1$ & $2$ & & & $-1$ & $1$ \\
$1$ & $\overline{1}$ & & $-1$ & $-1$ & $2$ &
$3$ & $1$ & & $-1$ & $1$ & $2$\\
$0$ & $\overline{1}$ & & & $1$ & $1$ &
$2$ & $1$ & & & $-1$ & $1$ \\
$\overline{1}$ & $\overline{1}$ & $1$ & & $-1$ & $2$ &
$1$ & $1$ & $1$ & & $1$ & $2$ \\  \hline
\multicolumn{12}{c}{} \\ \hline
\multicolumn{6}{c||}{$\Gamma=E_{1}'$} &
\multicolumn{6}{c}{$\Gamma=E_{2}'$} \\ \hline
$2\mu$ & $2\nu$ & $C_{1}$ & $C_{2}$ & $C_{3}$ & $N$ &
$2\mu$ & $2\nu$ & $C_{1}$ & $C_{2}$ & $C_{3}$ & $N$ \\ \hline
$1$ & $1$ & $i$ &  & $-\sqrt{2}$ & $3$ &
$3$ & $3$ & $i$ & & $-\sqrt{2}$ & $3$ \\
$\overline{1}$ & $1$ &  & $-1$ & $-\sqrt{2}$ & $3$ &
$\overline{3}$ & $3$ &  & $-1$ & $\sqrt{2}$ & $3$ \\
$1$ & $\overline{1}$ & & $-1$ & $-\sqrt{2}$ & $3$ &
$3$ & $\overline{3}$ & & $-1$ & $\sqrt{2}$ & $3$ \\
$\overline{1}$ & $\overline{1}$ & $i$ & & $\sqrt{2}$ & $3$ &
$\overline{3}$ & $\overline{3}$ & $i$ & & $\sqrt{2}$ & $3$ \\  \hline
\multicolumn{12}{c}{} \\ \hline
\multicolumn{12}{c}{$\Gamma=G'$} \\ \hline
$2\mu$ & $2\nu$ & $C_{1}$ & $C_{2}$ & $C_{3}$ & $N$ &
$2\mu$ & $2\nu$ & $C_{1}$ & $C_{2}$ & $C_{3}$ & $N$ \\ \hline
$3$ & $3$ & $i\sqrt{2}$ &  & $1$ & $3$ &
$3$ & $\overline{1}$ & & & $1$ & $1$ \\
$1$ & $3$ &  &  & $1$ & $1$ &
$1$ & $\overline{1}$ & & $\sqrt{2}$ & $-1$ & $3$ \\
$\overline{1}$ & $3$ & & & $1$ & $1$ &
$\overline{1}$ & $\overline{1}$ & $i\sqrt{2}$ & & $-1$ & $3$ \\
$\overline{3}$ & $3$ & & $\sqrt{2}$ & $1$ & $3$ &
$\overline{3}$ & $\overline{1}$ & & & $1$ & $1$ \\
$3$ & $1$ & & & $1$ & $1$ &
$3$ & $\overline{3}$ & & $\sqrt{2}$ & $1$ & $3$ \\
$1$ & $1$ & $i\sqrt{2}$ & & $1$ & $3$ &
$1$ & $\overline{3}$ & & & $-1$ & $1$ \\
$\overline{1}$ & $1$ & & $\sqrt{2}$ & $-1$ & $3$ &
$\overline{1}$ & $\overline{3}$ & & & $1$ & $1$ \\
$\overline{3}$ & $1$ & & & $1$ & $1$ &
$\overline{3}$ & $\overline{3}$ & $i\sqrt{2}$ & & $-1$ & $3$ \\ \hline
\end{tabular} 
}

\end{center}

\newpage
%\vspace{5mm}
\begin{center}

{\bf Table 3} $~~$ Character table of the tetrahedral double group {\bf T'}
$$\omega=\left(\overline{\omega}\right)^{-1}=\exp\{-i2\pi/3\}. $$

\vspace{2mm}
{\scriptsize
\begin{tabular}{c|ccccccc|c} \hline \hline 
& $E$ & $4C_{3}$ & $4C_{3}^{2}$ & $6C_{2}$ & $E'$ & $4C_{3}^{4}$ 
& $4C_{3}^{5}$ & $\mu$  \\  \hline
$A_{0}$ & $1$ & $1$ & $1$ & $1$ & $1$ & $1$ & $1$ & $0$ \\
$A_{+}$ & $1$ & $\omega$ & $\overline{\omega}$ & $1$ & $1$ & $\omega$ &
$\overline{\omega}$ & $0$ \\
$A_{-}$ & $1$ & $\overline{\omega}$ & $\omega$ & $1$ & $1$ 
& $\overline{\omega}$ & $\omega$ & $0$ \\
$T$ & $3$ & $0$ & $0$ & $-1$ & $3$ & $0$ & $0$ & $1,~0,~-1$ \\
$E_{0}'$ & $2$ & $1$ & $-1$ & $0$ & $-2$ & $-1$ & $1$ & $1/2,~-1/2$ \\
$E_{+}'$ & $2$ & $\omega$ & $-\overline{\omega}$ & $0$ & $-2$ & $-\omega$ &
$\overline{\omega}$ & $1/2,~-1/2$ \\
$E_{-}'$ & $2$ & $\overline{\omega}$ & $-\omega$ & $0$ & $-2$ 
& $-\overline{\omega}$ & $\omega$ & $1/2,~-1/2$ \\ \hline \hline
\end{tabular} 
}

\end{center}

\vspace{4mm}
\begin{center}

{\bf Table 4} $~~$ Irreducible bases in the group space of {\bf T'}
$$\begin{array}{c}
\psi_{\mu \nu}^{'\Gamma}=N^{-1/2} \displaystyle \sum_{i=1}^{4}~
C_{i}~\Phi_{\mu \nu}^{'(i)},\\
\tau=\exp\{-i\pi/4\},~~~~~
\omega=\left(\overline{\omega}\right)^{-1}=\exp\{-i2\pi/3\}. 
\end{array} $$
$$\begin{array}{l}
\psi_{00}^{A_{0}}=\left\{\Phi_{00}^{(1)}+\Phi_{00}^{(2)}
+\sqrt{2}\Phi_{00}^{(3)} +\sqrt{2}\Phi_{00}^{(4)}  \right\}/\sqrt{6}, \\
\psi_{00}^{A_{+}}=\left\{\Phi_{00}^{(1)}+\Phi_{00}^{(2)}
+\sqrt{2}\overline{\omega} \Phi_{00}^{(3)} 
+\sqrt{2} \omega \Phi_{00}^{(4)}  \right\}/\sqrt{6}, \\
\psi_{00}^{A_{-}}=\left\{\Phi_{00}^{(1)}+\Phi_{00}^{(2)}
+\sqrt{2} \omega \Phi_{00}^{(3)} 
+\sqrt{2} \overline{\omega} \Phi_{00}^{(4)}  \right\}/\sqrt{6}, \\
\end{array} $$

\vspace{2mm}
{\scriptsize
\begin{tabular}{c|c|ccccc||c|c|ccccc} \hline \hline 
\multicolumn{14}{c}{$\Gamma=T$}  \\ \hline
$\mu$ & $\nu$ & $C_{1}$ & $C_{2}$ & $C_{3}$ & $C_{4}$ & $N$ &
$\mu$ & $\nu$ & $C_{1}$ & $C_{2}$ & $C_{3}$ & $C_{4}$ & $N$ \\  \hline

$1$ & $1$ & $\sqrt{2}$ & & $i$ & $-i$ & $4$ &
$\overline{1}$ & $0$ & & & $1$ & $-i$ & $2$ \\
$0$ & $1$ &  &  & $i$ & $-1$ & $2$ &
$1$ & $\overline{1}$ & & $-\sqrt{2}$ & $-i$ & $-i$ & $4$ \\
$\overline{1}$ & $1$ & & $-\sqrt{2}$ & $i$ & $i$ & $4$ &
$0$ & $\overline{1}$ & & & $i$ & $1$ & $2$ \\
$1$ & $0$ & & & $-1$ & $-i$ & $2$ &
$\overline{1}$ & $\overline{1}$ & $\sqrt{2}$ & & $-i$ & $i$ & $4$ \\ 
\cline{8-14}
$0$ & $0$ & $1$ & $-1$ & & & $2$ &
\multicolumn{7}{c}{ } \\ \cline{1-7}
\multicolumn{14}{c}{} \\ \hline
\multicolumn{14}{c}{$\Gamma=E_{0}'$}  \\ \hline
$2\mu$ & $2\nu$ & $C_{1}$ & $C_{2}$ & $C_{3}$ & $C_{4}$ & $N$ &
$2\mu$ & $2\nu$ & $C_{1}$ & $C_{2}$ & $C_{3}$ & $C_{4}$ & $N$ \\ \hline
$1$ & $1$ & $\tau$ & & $1$ & $i$ & $3$ &
$1$ & $\overline{1}$ & & $\tau$ & $1$ & $1$ & $3$ \\
$\overline{1}$ & $1$ & & $i\tau$ & $1$ & $1$ & $3$ &
$\overline{1}$ & $\overline{1}$ & $-i\tau$ & & $1$ & $i$ & $3$ 
\\ \hline
\multicolumn{14}{c}{} \\ \hline
\multicolumn{7}{c||}{$\Gamma=E_{+}'$} &
\multicolumn{7}{c}{$\Gamma=E_{-}'$} \\  \hline
$2\mu$ & $2\nu$ & $C_{1}$ & $C_{2}$ & $C_{3}$ & $C_{4}$ & $N$ &
$2\mu$ & $2\nu$ & $C_{1}$ & $C_{2}$ & $C_{3}$ & $C_{4}$ & $N$ \\ \hline
$1$ & $1$ & $\tau$ & & $\overline{\omega}$ & $i\omega$ & $3$ &
$1$ & $1$ & $\tau$ & & $\omega$ & $i\overline{\omega}$ & $3$ \\
$\overline{1}$ & $1$ & & $i\tau$ & $\overline{\omega}$ 
& $\omega$ & $3$ &
$\overline{1}$ & $1$ & & $i\tau$ & $\omega$ 
& $\overline{\omega}$ & $3$ \\  
$1$ & $\overline{1}$ & & $i\tau$ & $i\overline{\omega}$ 
& $i \omega$ & $3$ &
$1$ & $\overline{1}$ & & $i\tau$ & $i \omega$ 
& $i\overline{\omega}$ & $3$ \\
$\overline{1}$  & $\overline{1}$ & $\tau$ & & $-i\overline{\omega}$ 
& $-\omega$ & $3$ &
$\overline{1}$ & $\overline{1}$ & $\tau$ & & $-i\omega$ 
& $-\overline{\omega}$ & $3$ \\  \hline
\end{tabular} 
}

\end{center}

\end{document}